\documentclass[letterpaper]{article}
\pdfoutput=1
\usepackage{amsthm,amsmath,amssymb}
\usepackage{color}
\usepackage{graphicx}
\usepackage{epsfig}
\usepackage{fancyhdr}
\usepackage{graphicx}        
\usepackage{multicol}        
\usepackage[bottom]{footmisc}
\usepackage{enumerate}
\usepackage{mathbbol,dsfont}
\usepackage{hyperref}

\usepackage{url}

\def\E{\mathds{E}}
\def\P{\mathds{P}}
\def\Q{\mathds{Q}}
\def\R{\mathds{R}}

\def\s{\mathfrak{s}}

\def\I{\mathcal{I}}

\def\d{\mathrm{d}}

\def\ds{\displaystyle}

\newtheorem{theorem}{Theorem}[section]
\newtheorem{lemma}[theorem]{Lemma}

\numberwithin{equation}{section}

\begin{document}

\title{Bridge Copula Model for Option Pricing}
\author{Giuseppe~Campolieti$^1$, Roman N. Makarov$^2$, and Andrey Vasiliev$^3$\\
Mathematics Department, Wilfrid Laurier University\\
75 University Avenue West, Waterloo, Ontario, Canada\\
E-mails: $^1$\texttt{gcampoli@wlu.ca},
$^2$\texttt{rmakarov@wlu.ca},  and $^3$\texttt{andrey.vas@gmail.com}}


\date{November 6, 2009}

\maketitle

\begin{abstract}
In this paper we present a new multi-asset pricing model, which is built upon
newly developed families of solvable multi-parameter single-asset diffusions with a nonlinear smile-shaped volatility and an
affine drift. Our multi-asset pricing model arises by employing copula methods. In particular,
all discounted single-asset price processes are modeled as martingale diffusions under a risk-neutral measure.
The price processes are so-called UOU diffusions and they are each generated by combining a variable (It\^{o}) transformation with a measure change performed on an underlying
Ornstein-Uhlenbeck (Gaussian) process. Consequently, we exploit the use of a normal bridge copula for coupling the single-asset dynamics
while reducing the distribution of the multi-asset price process to a multivariate normal distribution. Such an approach allows
us to simulate multidimensional price paths in a precise and fast manner and hence to price path-dependent financial
derivatives such as Asian-style and Bermudan options using the Monte
Carlo method. We also demonstrate how to successfully calibrate our
multi-asset pricing model by fitting respective equity option and asset market prices to the single-asset models
and their return correlations (i.e. the copula function) using the least-square and maximum-likelihood estimation methods.
\end{abstract}

\section*{Introduction}
Many quantitative finance applications require a
multi-asset pricing model with dependencies between the single-asset price components. Compared to the variety of univariate asset price models, the pool of
multi-asset pricing models is not so extensive. Most of multivariate models are based on multidimensional geometric Brownian
motion with the possible inclusion of a jump process.
In this paper we develop and explore a new multi-asset arbitrage-free pricing model based on a special family
of nonlinear diffusions. The development of efficient computational methods for pricing multi-asset equity derivatives under such a model and
the calibration of the multi-asset model to both standard equity option data as well as historical equity prices are the objectives of the current paper.

Here, we specialize on option pricing applications under so-called \textsf{UOU} diffusion models, which are obtained by transforming an underlying Ornstein-Uhlenbeck
diffusion process via the use of a diffusion canonical transformation method (see \cite{ACCL,AC05,CM06,CM08b} and references therein).
For all choices of model parameters, all discounted (single-asset) price processes \textsf{UOU} are conservative martingales
under a risk-neutral measure. Since the univariate diffusions are solvable, the single-asset risk-neutral transition probability density function is
given in analytically closed form. Moreover, implied volatility surfaces for
this highly nonlinear asset price model exhibit a wide range of pronounced smiles and
skews of the type observed in the option markets. The main relevant features of the univariate \textsf{UOU} model are summarized in Section~\ref{Section1}.

To construct a multivariate probability distribution, one can use a copula function 
that allows us to couple univariate distribution functions.
Sampling from the obtained joint multivariate distribution function thereby reduces to sampling from the copula function and from the univariate distributions.
Therefore, the copula method allows us to construct the joint distribution and density functions as well as to obtain an exact path sampling algorithm.

The main computational disadvantage of such an approach is the calculation of inverses of the distribution functions. This operation can be a
rather time-consuming computational problem for a~complicated multi-parameter distribution. Nevertheless, such a drawback can be significantly
improved if the copula function and univariate
distributions have a similar structure. As is shown in
Subsection~\ref{subsect1.3}, the bridge probability density function
(conditional on the values of the process at the endpoints of a time
interval) of a \textsf{UOU} diffusion is reduced to a normal density. Hence it is natural to couple univariate \textsf{UOU} bridges using a Gaussian copula.
Based on this idea, in Section~\ref{sect2}, we construct a two-step algorithm for the \textit{exact} path simulation of the multidimensional (nonlinear) \textsf{UOU} process in the risk-neutral measure. Firstly, we
apply a usual copula method for sampling the multi-asset process at the terminal (maturity) time. Secondly,
we use a bridge sampling along with a multivariate normal distribution to model the  process at any intermediate time.

In Section~\ref{sect3}, we demonstrate the calibration of the univariate and multivariate models to historical asset and equity option prices. The
calibration process has two stages. First, we calibrate all univariate (marginal) asset price models independently of each other.
Using the least-square method the models can be fitted to standard European option prices. Alternatively, the maximum likelihood estimation (MLE)
allows the models to be fitted to historical asset prices. Second, we fit the copula function to historical observations. Since, our model
assumes a normal copula, we need to find a best-fitted normal correlation matrix.

In Section~\ref{sect4}, we give computational applications of the model to pricing multi-asset path-dependent Asian-style and Bermudan options. In pricing Bermudan options we use a regression-based Monte Carlo method.

In summary, the main results of our paper include: the construction a new family of multivariate models
for which marginal processes are local volatility smile diffusions; the development of calibration schemes for the single-asset
and multi-asset pricing \textsf{UOU} diffusion models based on the least-square and MLE methods;
the construction of an exact multivariate path simulation method that
can be used for Monte Carlo pricing of generally path-dependent European-style and American-style options.

\section{Ornstein-Uhlenbeck Family of Univariate State-Dependent Vo\-latility Diffusion Models} \label{Section1}

\subsection{Diffusion Canonical Transformation} \label{subsect1.1}

The diffusion canonical transformation method, first presented in
\cite{ACCL} and then further developed in \cite{Kuzn,CM06,CM08b}, leads to
various families of solvable one-dimensional diffusions with a nonlinear diffusion coefficient function and an affine drift. In this paper, we consider the $\mathsf{UOU}$ family, which is based on a regular Ornstein-Uhlenbeck
process $(X_t)_{t\geq 0}\in\I\equiv\R$. The regular Ornstein-Uhlenbeck process is defined by the infinitesimal generator
\begin{equation}
 \mathcal L f(x)\equiv \frac{1}{2}\nu^2 f^{\prime\prime}(x)
 -\lambda x f^\prime(x)\,,\;\;
  x \in \R, \label{XGenerator}
\end{equation}
where $\lambda>0$ and $\nu>0$ are constants. Both left and right boundaries $l=-\infty$ and $r=\infty$ of the state space $\I$
are non-attracting natural for all choices of parameters. The transition
probability density function (PDF) is
\begin{equation} \label{OUden}
 p_X(t;x_0,x)=\sqrt{\frac{\kappa}{2\pi(1-e^{-2\lambda t})}}
 \exp\left(-\frac{\kappa(x-x_0 e^{-\lambda t})^2}{2(1-e^{-2\lambda
 t})}\right)\,,
\end{equation}
where we define $\kappa \equiv \frac{2\lambda}{\nu^2}>0$.

Let $\rho > 0$ be a strictly positive constant. Then, two \textit{fundamental solutions} of $\mathcal{L} \varphi (x)
= s \varphi(x)$, $x\in\I$, are given by
\begin{equation} \label{OUfundsol}
\varphi^-_\rho(x)=\exp\left(\frac{\kappa
x^2}{4}\right)D_{-\upsilon}(x\sqrt{\kappa})\quad\mbox{and}\quad
\varphi^+_\rho(x)=\varphi^-_\rho(-x), \end{equation}
where $\upsilon \equiv \rho/\lambda>0$ and $D_{-\upsilon}(\cdot)$ is Whittaker's
parabolic cylinder function (see \cite{AS72}).
The solutions $\varphi^+_\rho(x)$ and $\varphi^-_\rho(x)$ are linearly independent and respectively
increasing and decreasing positive functions of $x\in \R$.

We now construct another diffusion process $(S_t)_{t\geq 0}\in \mathcal{D}=\R_+=(0,\infty)$
by applying a diffusion canonical transformation to the underlying Ornstein-Uhlenbeck diffusion.
This process obeys the stochastic differential equation (SDE)
\begin{equation} \label{FSDE}
 \d S_t=rS_t\d t+\sigma(S_t)\d W_t, \,\,S_{t=0}=S_0\,,
\end{equation}
where $r$ is a constant and $\sigma(S)$ is a nonlinear diffusion coefficient function. Here $({W}_t)_{t\ge 0}$ is
a standard Brownian motion in some appropriate probability measure~${\P}$.

The initial step of the transformation is to apply a Doob's $h$-transform or $\rho$-excessive transform (see
\cite{BS02}) to $(X_t)$. The resulting diffusion process
$\big(X^{(\rho)}_t\big)_{t\geq 0}$ is defined by the following
infinitesimal generator:
\begin{equation} \label{XrhoGen}
\mathcal{L}^{(\rho)} f(x)=\frac{1}{2}\nu^2f^{\prime\prime}(x)
+\lambda_\rho(x)f^{\prime}(x),\; x\in\I,
\end{equation}
where $\lambda_\rho(x)\equiv -\lambda x+\nu^2\hat{u}_\rho'(x)/\hat{u}_\rho(x)$ for any $\rho>0$. The
strictly positive {\it generating function} is given by
$\hat{u}_\rho(x)=q_1 \varphi^+_\rho(x) + q_2 \varphi^-_\rho(x)$ with
parameters $q_1,q_2\geq 0,$ $q_1+q_2>0$. The process $(X^{(\rho)}_t)$ has the
transition PDF
\begin{equation} p_X^{(\rho)}(t;x_0,x)=e^{-\rho t}\frac{\hat
u_\rho(x)}{\hat u_\rho(x_0)}
   p_X\left(t;x_0,x\right),\; x,x_0\in\I,\,t>0. \label{urho}
\end{equation}


The final step of the transformation is a change of variable (see \cite{CM08b} for a more general discussion). We define a new
process $S_t=\mathsf{F}(X_t^{(\rho)})$, $t\geq 0$, that solves SDE (\ref{FSDE}) by
finding a strictly monotonic map that solves $\mathcal{L}^{(\rho)}\mathsf{F}(x) = r \mathsf{F}(x)$, for constant $r$. Then
$\sigma(\mathsf{F}(x)) = \nu \vert\mathsf{F}'(x)\vert$ or equivalently $\sigma(s) = \nu/\vert\mathsf{X}'(s)\vert$,
where $\mathsf{X}\equiv \mathsf{F}^{-1}$ defines the unique inverse map. The transition PDF
$p_S$ for the process $(S_t)$ follows from that for the underlying process $(X_t)$:
\begin{equation}\label{PrKernelF}
  p_S(t;s_0,s)=\frac{\nu}{\sigma(s)}\;p_X^{(\rho)}(t;x_0,x)=
  \frac{\nu}{\sigma(s)}\frac{\hat u_\rho(x)}{\hat u_\rho(x_0)}
  e^{-\rho t} p_X\left(t;x_0,x\right),
\end{equation}
$x=\mathsf{X}(s)$, $x_0=\mathsf{X}(s_0)$.

We now apply the above construction to a subfamily of diffusions with choice $q_1=0, q_2=1$, i.e. with generating function
$\hat{u}_\rho(x)=\varphi^-_\rho(x)$. Letting $r+\rho > 0$ and $\rho > 0$, we specifically consider
\begin{equation} \label{FmapUOU}
\mathsf{F}(x)=c\frac{\varphi^+_{r+\rho}(x)}{\varphi^-_\rho(x)}= c\frac{D_{- \upsilon -r/\lambda}(-x\sqrt{\kappa})}{D_{-\upsilon}(x\sqrt{\kappa})}\,.
\end{equation}
This function maps $x\in\R$ onto
$s\in(0,\infty)$ and is monotonically increasing, where $s=\mathsf{F}(x)$
has the unique inverse relation $x=\mathsf{X}(s)$. This
transformation leads to a family of diffusions $(S_t)$ that is referred to as the
unbounded Ornstein-Uhlenbeck ($\mathsf{UOU}$) family with the
diffusion coefficient function given by
 \begin{equation} \label{SigmaUOU}
 \sigma(s) = \nu\sqrt{\kappa}e^{\frac{\kappa x^2}{4}}\left(
  (\upsilon+\frac{r}{\lambda})
  \frac{D_{-\upsilon -1 - \frac{r}{\lambda}}(-\sqrt{\kappa}x)}{D_{-\upsilon}(\sqrt{\kappa}x)}
  + \upsilon\frac{D_{-\upsilon-\frac{r}{\lambda}}(-\sqrt{\kappa}x)}{D_{-\upsilon}(\sqrt{\kappa}x)}
  \frac{D_{-\upsilon-1}(-\sqrt{\kappa}x)}{D_{-\upsilon}(\sqrt{\kappa}x)}
 \right),
 \end{equation}
where $x=\mathsf{X}(s)$. The volatility function $\sigma$ in (\ref{SigmaUOU}) depends on several
adjustable positive parameters such as $c, \nu, \lambda, \rho$, and more generally $r\ne 0$
also enters as an additional parameter within the volatility specification. Notice that for the driftless case with
$r=0$ formula (\ref{SigmaUOU}) simplifies as follows:
$\sigma(\mathsf{F}(x)) =
\ds\frac{\sigma_0\s(x)}{\hat{u}_\rho^2(x)},$ where
$\sigma_0>0$ is a constant and $\s(x)=e^{\frac{\kappa x^2}{2}}$ is the scale density
for the $X$-diffusion.

The following is an important statement for the purposes of risk-neutral pricing.

\begin{lemma}[\cite{CM08b}] \label{lemma1} Consider a process $(S_t)_{t\ge 0}\in \R_+$ of the \textsf{UOU} family solving the SDE
(\ref{FSDE}) with the diffusion coefficient $\sigma$ specified by
(\ref{SigmaUOU}) and having transition PDF $p_S$ specified by equations (\ref{PrKernelF}), (\ref{FmapUOU}), (\ref{SigmaUOU}) and given by \begin{equation} \label{PDF_UOU}
p_S(t;s_0,s)= \frac{e^{-\rho
t+\kappa(x^2-x_0^2)/4}}{c\mathcal{W}(x)}\frac{D_{-\upsilon}^{\,3}(x\sqrt{\kappa})}{
D_{-\upsilon}(x_0\sqrt{\kappa})} p_X(t;x_0,x),\quad t>0,\;s,s_0>0,
\end{equation}
where $\mathcal{W}(x)=W[D_{-\upsilon}(x\sqrt{\kappa}),D_{-\upsilon-r/\lambda}(-x\sqrt{\kappa})]>0$ is
the Wronskian\footnote{The Wronskain $W$ of functions $f$ and $g$ is defined by $W[f(x),g(x)]=f(x)g'(x) - f'(x)g(x)$.}, and
$x = \mathsf{X}(s)$, $x_0=\mathsf{X}(s_0)$. Then $(S_t)$ is a conservative process, i.e. $\P(S_t\in\R_+)=1$ for all $t>0$.
Moreover, the discounted asset price process $(e^{-rt}S_t)_{t\geq 0}$ is a true martingale.
\end{lemma}

\subsection{Pricing Vanilla Options} \label{subsect1.2}

According to Lemma~\ref{lemma1} there exists an equivalent martingale measure under the \textsf{UOU} family with transition PDF (\ref{PDF_UOU})
for any choice of the model parameters. Consider a standard European-style option defined by its payoff function $\Lambda(S)$ at
terminal price $S=S_T$ and maturity (expiration) time $T>t_0=0$. For example, a vanilla European call has the payoff function $C^{E}(S)=(S-K)^+ \equiv
\max\{S-K,0\},$ where $K>0$ is a strike price. The valuation of a
standard European option is given by the conditional expectation under a risk-neutral probability measure
$\P\equiv \Q$:
\begin{equation} \label{EurOpt}
 V(S_0,T)=e^{-r T}\E^\Q[\Lambda(S_T)\mid S_{t=0} = S_0] = e^{-r T}\E^\Q[\Lambda(\mathsf{F}(X_T^{(\rho)}))\mid
 X_0^{(\rho)}=x_0]\,.
\end{equation}
This is reduced to the valuation of a one-dimensional integral
expressible as follows:
\begin{equation}  \label{EurOptFX} \begin{array}{rcl} V(S_0,T) &=& e^{-r  T}\ds\int_0^\infty
p_S(T;S_0,s)\Lambda(s)\;\d s\\[10pt] &=&
\ds\frac{e^{-(r+\rho)T}}{\hat{u}_\rho(x_0)}\int_{-\infty}^\infty
\hat{u}_\rho(x)\,p_X(T;x_0,x) \Lambda(\mathsf{F}(x))\,\d x\,,
\end{array}
\end{equation}
where $x_0=\mathsf{X}(S_0).$

\begin{figure}
  \centering
  \includegraphics[width=0.4\linewidth]{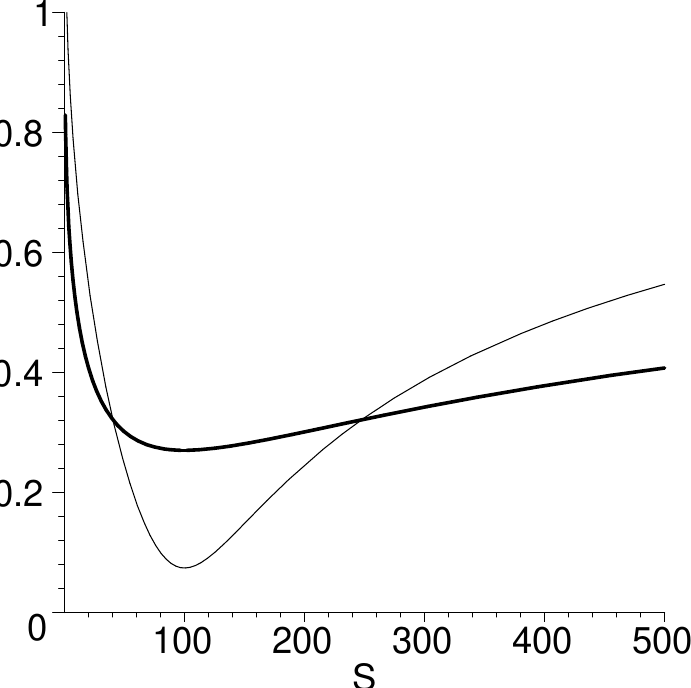}\hfill\includegraphics[width=0.5\linewidth]{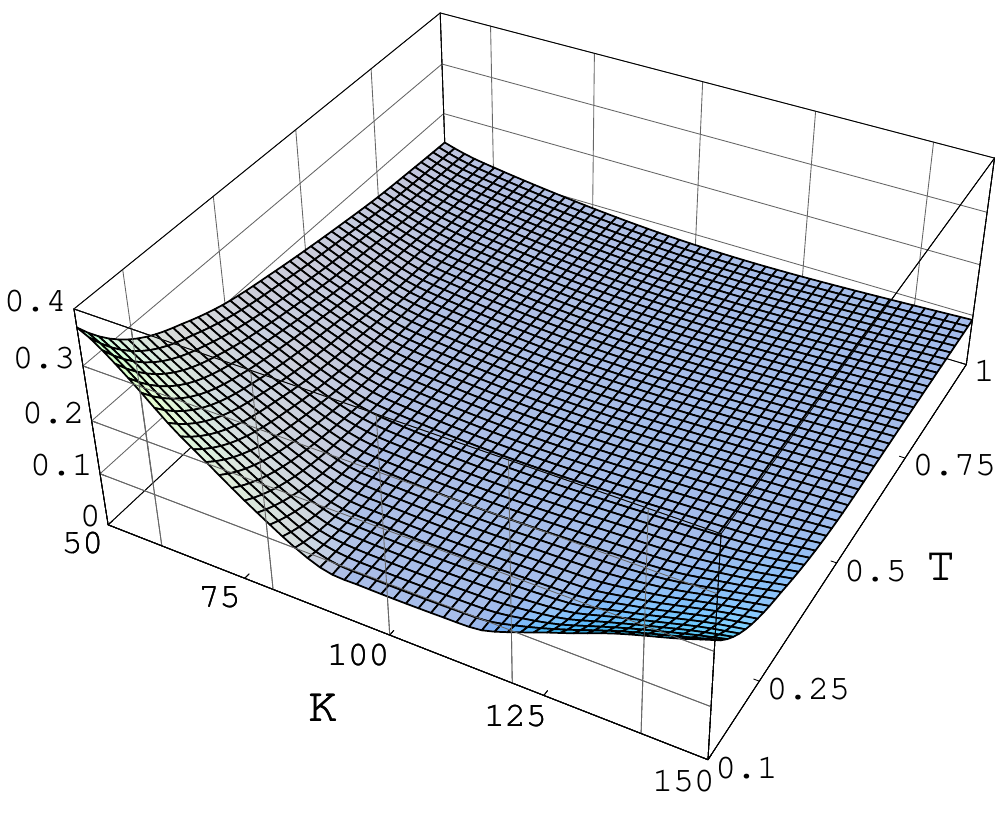}
\vspace*{8pt}
  \caption{ Local volatility (or log-normal volatility) function $\sigma(S)/S$
  (left plot) and Black-Scholes impled volatility surface (right plot)
  for the $\mathsf{UOU}$ family.
  The $\mathsf{UOU}$ family is plotted using model parameters $c=100$;
  $\rho=0.005$, $\kappa=5$, $\upsilon=0.1$ (thin line) and
  $\rho=0.02$, $\kappa=1$, $\upsilon=0.5$ (thick line).
  The implied volatility surface, with $T$ as time to maturity and $K$ as strike price, corresponds to the same
choice of parameters as the local volatility plot drawn with the thin line.
  The interest rate $r=0.1$ and spot $S_0=100$ are used.}
  \label{Fig1}
\end{figure}

\subsection{The \textsf{UOU} Bridge} \label{subsect1.3}

Let us consider a bridge process on $[t_1,t_2]$, $0\leq t_1<t_2$, generated by a single asset price process $S_t$ with $S_{t_1}$ and
$S_{t_2}$ fixed at $s_1$ and $s_2$, respectively. The bridge
density $b_S(t;s)\equiv b_S(t_1,t_2,t;s_1,s_2,s)$ for $S_t = s$,
$t_1<t<t_2$, conditional on $S_{t_1}=s_1$ and
$S_{t_2}=s_2$, is given by
\begin{equation} \label{UBrdgDen}
 \begin{split}
  &b_S(t;s)\equiv \displaystyle \frac{p_S(t-t_1;s_1,s)\,p_S(t_2-t;s,s_2)}{
   p_S(t_2-t_1;s_1,s_2)}\\[4mm]
   \mbox{} &= \displaystyle\frac{\nu}{\sigma(s)}\,
     \frac{p_X^{(\rho)}(t-t_1;x_1,\mathsf{X}(s))\,
   p_X^{(\rho)}(t_2-t;\mathsf{X}(s),x_2)}{p_X^{(\rho)}(t_2-t_1;x_1,x_2)}\\[5mm]
   \mbox{} &= \displaystyle \frac{\nu}{\sigma(s)}\,
     \frac{p_X(t-t_1;x_1,\mathsf{X}(s))\,
   p_X(t_2-t;\mathsf{X}(s),x_2)}{p_X(t_2-t_1;x_1,x_2)}
   \equiv \displaystyle\frac{\nu}{\sigma(s)}b_X(t;\mathsf{X}(s)),
 \end{split}
\end{equation}
where $x_1\equiv\mathsf{X}(s_1)$, $x_2\equiv\mathsf{X}(s_2)$,  and $b_X(t;x)\equiv b_X(t_1,t_2,t;x_1,x_2,x)$ is the bridge PDF of the underlying $X$-diffusion
conditional on the endpoint values $X_{t_1}=x_1$ and $X_{t_2}=x_2$.
Since the underlying process
$(X_t)$ is a Gaussian process, the bridge
process, generated by $(S_t)_{t\geq 0}$, with transition
PDF (\ref{UBrdgDen}) is just a nonlinear transformation of a
Gaussian bridge, with respective path points mapped as $X_t = \mathsf{X}(S_t)$.
Thus, the bridge PDF $b_S$ for $(S_t)$ from the
\textsf{UOU}~family can be reduced to a Gaussian PDF for the underlying $X$-process. Indeed, from Eq.~(\ref{OUden}) we see that the bridge PDF $b_X$ of the Ornstein-Uhlenbeck diffusion is a normal density $\varphi(x)=\frac{1}{\sqrt{2\pi} b}e^{-(x-\mu)^2/2b^2}$
with mean $\mu$ and variance $b^2$ given by
\begin{equation} \label{eqn4ab} \begin{array}{rcl}
   \mu &=& \displaystyle\frac{x_1 e^{\lambda \Delta_1} (e^{2\lambda \Delta_2}-1)
          +x_2 e^{\lambda \Delta_2} (e^{2\lambda \Delta_1}-1)}{e^{2\lambda
   (\Delta_1+\Delta_2)}-1},\\[10pt]
   b^2 &=& \displaystyle\frac{(e^{2\lambda \Delta_1}-1)(e^{2\lambda \Delta_2}-1)}{\kappa(e^{2\lambda
   (\Delta_1+\Delta_2)}-1)}, \end{array}
\end{equation}
where $\Delta_1\equiv t-t_1,$ $\Delta_2\equiv t_2-t$, and $\Delta_1
+ \Delta_2 = t_2-t_1$.

\section{Multivariate \textsf{UOU}-Diffusion Pricing Model} \label{sect2}

\subsection{Coupling \textsf{UOU} Processes} \label{subsect2.1}

Our goal  is to construct a multi-asset price process $(\mathbf{S}_t)_{t\geq 0}$ with
$\mathbf{S}_t\equiv \left(S_t^{1},\ldots,S_t^{n}\right)$, where each individual asset price process $(S_t^{k})_{t\geq 0}$,
$k=1,2,\ldots,n,$ is a univariate \textsf{UOU} diffusion obeying (\ref{FSDE}) with common drift parameter $r$ and diffusion function $\sigma = \sigma^k$.
Suppose that each of the $n$ univariate processes is described by
its own set of positive parameters $\xi_k=\left\{\lambda_k,\nu_k,c_k,\rho_k\right\},$
$k=1,2,\ldots,n.$ We denote by $p^k_{S}$ and $\sigma^k$ the univariate risk-neutral
transition PDF of the form (\ref{PDF_UOU}) and the diffusion coefficient given by (\ref{SigmaUOU}), respectively, which both correspond to the
$k$-th asset price process. In the following, for each $k=1,2,\ldots,n$,  $\hat{u}^k_{\rho_k}$, $\mathsf{F}^k$, and $\mathsf{X}^k$ will respectively
denote the generating function, the mapping function, the inverse mapping function of the $k$th diffusion model. The transition PDFs $p^k_{X}$ and $p_{X}^{(\rho_k,k)}$
correspond to the underlying diffusion $(X^k_t)_{t\geq 0}$ and the transformed diffusion $\left(X_t^{(\rho_k,k)}\right)_{t\geq 0}$, respectively.
All the above functions indexed by $k$ are obtained by using the parameters from $\xi_k$ in the respective equations provided in Section~\ref{Section1}.

Recall that the processes $(S^k_t)_{t\geq 0}$, $k=1,2,\ldots,n$, are defined by the
transformation $\mathsf{F}^k$ given by (\ref{FmapUOU}), i.e. $S^k_t = \mathsf{F}^k(X_t^{(\rho_k,k)})$ where $(X_t^{(\rho_k,k)})_{t\geq
0}$ are defined by (\ref{XrhoGen}) with $\lambda=\lambda_k$, $\nu=\nu_k$, and $\rho=\rho_k$. For an infinitesimal time increment $\delta t$, consider $\delta S^k_t\equiv S^k_{t+\delta t} - S^k_t$ and $\delta X_t^{(\rho_k,k)}\equiv X_{t+\delta t}^{(\rho_k,k)} - X_t^{(\rho_k,k)}$. It
follows that $\delta S^k_t \simeq (\mathsf{F}^k)'(X_t^{(\rho_k,k)})\cdot\delta X_t^{(\rho_k,k)}$.
From the map, $(\mathsf{F}^k)'(x) = \sigma^k(\mathsf{F}^k(x))/\nu_k$, we readily see that
the respective correlations between the ratios of the infinitesimal increment
and the diffusion coefficient remain the same after the change of variables:
\begin{equation} \label{ConsCorr}
 \mathrm{corr\,}\left(
 \frac{\delta S_t^{k}}{\sigma^k(S_t^{k})},\frac{\delta S_t^{\ell}}{\sigma^\ell(S_t^{\ell})}\right)
 \simeq \mathrm{corr\,}\left(
 \frac{\delta X_t^{(\rho_k,k
 )}}{\nu_k},\frac{\delta X_t^{(\rho_\ell,\ell)}}{\nu_\ell}\right),\quad k,\ell=1,\ldots,n.
\end{equation}

This relation shows how to couple \textsf{UOU} diffusions. First of all, we can
directly couple the processes $\left(X_t^{(\rho_k,k)}\right)$, which, in turn, introduces correlations among the asset price processes $(S^k_t)$.
In fact, the same method is used to couple geometric Brownian
motions that are just exponentially transformed
Brownian motions with drift. Hence, by using dependent Brownian motions,
one introduces the correlations between the log-returns. In the case of more general
diffusions, one may introduce correlations between the
volatility-scaled returns. Alternatively, one may
couple the underlying $X$-processes.
For example, it is possible to use a multivariate Ornstein-Uhlenbeck process as an~underlying vector process.
Another approach (which is not discussed in this paper) is to extend the diffusion canonical transformation
method directly to the case of multivariate diffusion processes.

Another idea, and the one that we implement in this paper, is the coupling of $X^{(\rho)}$-bridges.
As is shown in Subsection~\ref{subsect1.3}, the bridge PDF of a \textsf{UOU} $X^{(\rho)}$-process
is the same as the density function of the Ornstein-Uhlenbeck bridge, which is a multivariate normal PDF.
The respective multivariate distribution function obtained with the use of a Gaussian copula is nothing more than a multivariate normal CDF. Therefore, the coupling of $X^{(\rho)}$-bridges considerably simplifies the form of the joint path distribution function as well as the path simulation algorithm.

\subsection{Copula Function} \label{subsect2.2}
A copula $\mathcal{C}(u_1,u_2,\ldots,u_n)$ is a multivariate
cumulative distribution function (CDF) defined on the
$n$-dimensional unit cube $[0, 1]^n$ such that every marginal
distribution is uniform on the unit interval $[0, 1]$ (for a more
detailed definition we refer to \cite{Nel99}). Suppose that
$\Phi^1,\Phi^2,\ldots,\Phi^n$ are univariate distribution functions,
e.g., $\Phi^k(x)=\int_{-\infty}^x f^k(y)\,\d y$, where $f^k$ is a respective
univariate PDF. It follows that $\mathcal{C}(\Phi^1(x_1),\ldots,\Phi^n(x_n))$ is a
multivariate CDF with marginals
$\Phi^k(x)=\mathcal{C}(1,\ldots,1,u_k=\Phi^k(x),1,\ldots,1),$
$k=1,2,\ldots,n$. The well-known Sklar's theorem states that any
$n$-dimensional joint distribution function $\mathbf{\Phi}$ with
continuous marginals $\Phi^1,\ldots,\Phi^n$ has a unique copula
representation:
\begin{equation} \label{multvarcdf}
\mathbf{\Phi}(x_1,x_2,\ldots,x_n)=\mathcal{C}(\Phi^1(x_1),\Phi^2(x_2),\ldots,\Phi^n(x_n)).
\end{equation} In other words, for continuous multivariate distributions,
the marginal distributions and the multivariate
dependence structure can be separated.  The multivariate density
function $\mathbf{f}$ is then obtained by differentiating
Eq.~(\ref{multvarcdf}): \begin{equation} \label{copden}
\mathbf{f}(x_1,\ldots,x_n)=\frac{\partial^n
\mathcal{C}(\Phi^1(x_1),\ldots,\Phi^n(x_n))}{
\partial u_1\ldots\partial u_n} f^1(x_1) \ldots
f^n(x_n).\end{equation}

As a corollary of Sklar's theorem we have a class of copula
functions constructed from continuous multivariate probability distributions as follows:
\begin{equation} \label{multvarcop}
\mathcal{C}(u_1,u_2,\ldots,u_n)=\mathbf{\Phi}\left((\Phi^1)^{-1}(u_1),(\Phi^2)^{-1}(u_2),\ldots,(\Phi^n)^{-1}(u_n)\right)\,,
\end{equation}
for any $(u_1,u_2,\ldots,u_n)\in [0,1]^n$, where
$(\Phi^k)^{-1}$ is the inverse of~$\Phi^k$.

The Gaussian (or normal) copula is one of the most important in
financial applications. This copula is constructed
from the multivariate normal distribution:
\begin{equation} \label{normalcop}
\mathcal{C}^{Gauss}_R(u_1,u_2,\ldots,u_n)=\mathcal{N}_R\left(\mathcal{N}^{-1}(u_1),\mathcal{N}^{-1}(u_2),\ldots,\mathcal{N}^{-1}(u_n)\right)\,,
\end{equation}
where $\mathcal{N}_R$ is the standard $n$-variate normal CDF with mean vector zero, unit variances, and correlation matrix $R$.
Here, $\mathcal{N}^{-1}$ stands for the inverse of a standard univariate Gaussian CDF.

Consider the problem of sampling from a multivariate distribution
given by (\ref{multvarcdf}). The modelling algorithm consists of two
steps. First, we simulate a uniformly distributed vector $(U_1,U_2,\ldots,U_n) \in [0,1]^n$ from the copula
$\mathcal{C}$. After that, we sample a vector
$(X_1,X_2,\ldots,X_n)$ from the marginal distributions by evaluating the inverse CDFs:
$X_k=(\Phi^k)^{-1}(U_k),$ $k=1,2,\ldots,n$.
As is seen, the crucial part of this algorithm is the efficient computation of the inverse of a distribution function.

\subsection{Multivariate Path Copula} \label{subsect2.3}

The objective of this subsection is the derivation of a multivariate path distribution function in closed form. Such a function allows us to obtain an exact path simulation method and also to construct the joint path density function, which is used for calibrating the model with the maximum likelihood estimation method.

We employ the copula method to construct the joint distribution function of the multivariate process
$\mathbf{X}_t^{(\boldsymbol\rho)}=\big(X_{t}^{(\rho_1,1)},\ldots,X_{t}^{(\rho_n,n)}\big).$
The multi-asset price process $\mathbf{S}_t=\big(S_{t}^1,\ldots S_{t}^{n}\big)$ is then obtained by applying the respective
mapping function to each univariate $X^{(\rho)}$-diffusion: $S_{t}^k = \mathsf{F}^k\left(X_{t}^{(\rho_k,k)}\right)$, $k=1,2,\ldots,n$.

Suppose that the process $(\mathbf{X}_t^{(\boldsymbol\rho)})$ conditional on $\mathbf{X}_0^{(\boldsymbol\rho)}$ is to be sampled at a set of
times $\mathbf{T}=\{t_j\}_{j=1}^N\in [0,T]$, $T>0$, so that $0=t_0<t_1<t_2<\cdots<t_N=T$. Let $\widetilde{\mathbf{T}}=\{\tilde{t}_j\}_{j=1}^N$
represent some arrangement of time points in $\mathbf{T}$.
The ordering of the time points is determined by the simulation method used.
For the (forward) sequential method we assume that $0<\tilde{t}_1<\cdots<\tilde{t}_N=T$,
i.e. $\forall j\geq0\;\tilde{t}_j=t_j$. For the backward-in-time bridge method we have
that $0<\tilde{t}_N<\tilde{t}_{N-1}<\cdots<\tilde{t}_2<\tilde{t}_1=T$, i.e. $\forall j\geq 1\;\tilde{t}_{j}=t_{N+1-j}$.
In other words, we first obtain the value of the process at the terminal time $T$ and then simulate a bridge path conditionally
on the previously sampled value and the initial value. Notice that for the bridge sampling method, the ordering of intermediate
time points $\tilde{t}_2,\ldots,\tilde{t}_N$ may vary, e.g. one can use a forward bridge sampling with $0<\tilde{t}_2<\tilde{t}_3<\cdots<\tilde{t}_N<\tilde{t}_1=T$ or
a full bridge sampling method, where we successively halve the time interval or its largest segment to sample the
process at the midpoint of the time segment conditionally on the previously
sampled values at the endpoints of the current time (sub-)interval. The full bridge sampling may be applied along with the quasi-Monte Carlo method.

Let $f^k_j(x)$ denote the PDF of $X_{\tilde{t}_j}^{(\rho_k,k)}$ conditional on the $\sigma$-algebra $\mathcal{F}^k_{j-1}$
generated by the previously sampled $j$ path points $X_{0}^{(\rho_k,k)},X_{\tilde{t}_1}^{(\rho_k,k)},\ldots,X_{\tilde{t}_{j-1}}^{(\rho_k,k)}$,
where $1\leq j\leq N$ and $1\leq k\leq n$. For the sequential path sampling method, with $\tilde{t}_i=t_i$, we have that
$f^k_j(x) = p_X^{(\rho_k,k)}\left( t_j-t_{j-1}; X_{t_{j-1}}^{(\rho_k,k)}, x\right).$

For the backward bridge method we have that
$f^k_1(x) = p_X^{(\rho_k,k)}\left( T; X_{0}^{(\rho_k,k)},x\right)$.
Since all univariate processes are Markovian and are sampled backward in time, for  each $j=2,3,\ldots,N$,
$f^k_j(x)$ is a bridge PDF $b_X^k\left( \tilde{t}_j; x\right)$ of the Ornstein-Uhlenbeck bridge conditional on $X_{0}^{(\rho_k,k)}$ and $X_{\tilde{t}_{j-1}}^{(\rho_k,k)}$, where $\tilde{t}_{j}=t_{N+1-j}$ and $\tilde{t}_{j-1}=t_{N+2-j}$.
As is shown in Subsection 1.3, the PDF $b_X^k$ is a normal density function with mean $\mu_{kj}$ and variance $b^2_{kj}$,
with values given by (\ref{eqn4ab}) where $\lambda=\lambda_k,$ $\kappa=\kappa_k$, $\Delta_1=\tilde{t}_j-t_{0}=\tilde{t}_j,$ and $\Delta_2=\tilde{t}_{j-1}-\tilde{t}_{j}.$

In $X$-space, the joint CDF $\mathbf{\Phi}_j$ and respective joint PDF $\mathbf{f}_j$ of
the $n$-dimensional point $\mathbf{X}_{\tilde{t}_j}^{(\boldsymbol\rho)}$, $j=1,2,\ldots,N$, are then constructed by employing equations
(\ref{multvarcdf}) and (\ref{copden}), respectively, where the
marginal distribution functions are $\Phi^k_j(x)=\int_{-\infty}^x
f^k_j(x')\;\d x'$, $k=1,2,\ldots,n$, $j=1,2,\ldots,N$, $x\in\R$, $t>0$.
Notice that, for the backward-in-time bridge method, the CDF $\Phi^k_j(x)$ is a normal CDF $\mathcal{N}\left(\frac{x-\mu_{kj}}{b_{kj}}\right)$, for each $j=2,\ldots,N$.

From the Markov property of process $(\mathbf{X}_t^{(\boldsymbol\rho)})$, the multivariate path distribution function
$\mathbb{\Phi}$ and the respective joint path density function $\mathbb{f}$ of
$\left( \mathbf{X}^{(\boldsymbol\rho)}_{\tilde{t}_j}\right)_{j=1,\ldots,N}$
conditional on $\mathbf{X}^{(\boldsymbol\rho)}_0$ are given by the respective products
\begin{equation}
  \label{pathDF}
  \mathbb{\Phi}(\mathbf{x}_1,\ldots,\mathbf{x}_N) = \prod_{j=1}^N \mathbf{\Phi}_j(\mathbf{x}_j) \text{ \ and \ }
  \mathbb{f}(\mathbf{x}_1,\ldots,\mathbf{x}_N) = \prod_{j=1}^N \mathbf{f}_j(\mathbf{x}_j),
\end{equation}
where $\mathbf{x}_j\equiv(x_j^1,\ldots,x_j^n),$ $j=1,2\ldots,N$. In $S$-space, the joint path PDF $\mathbb{f}^s$ and the joint PDFs $\mathbf{f}^s_j$ of $\mathbf{S}_{\tilde{t}_j}$, $j=1,2\ldots,N$, are respectively given by
\begin{equation} \label{jointpathSPDF}
 \mathbb{f}^s(\mathbf{s}_1,\ldots,\mathbf{s}_N) = \prod_{j=1}^N \mathbf{f}^s_j(\mathbf{s}_j),\quad \mathbf{f}^s_j(\mathbf{s}_j) = \prod_{k=1}^n \frac{\nu_k}{\sigma^k(s^k_j)} f_j^k\left(\mathsf{X}^k(s_j^k) \right),
\end{equation}
where $\mathbf{s}_j=(s_j^1,\ldots,s_j^n)$

To sample from the distribution function in (\ref{pathDF}) we need a fast and
accurate algorithm for inverting univariate CDFs. Usually, for
non-standard distributions such as the PDF given by formula
(\ref{PDF_UOU}), such algorithms are quite computationally
intensive. Therefore, the sequential approach becomes rather unfeasible
for the path simulation of the multi-asset price \textsf{UOU} process $(\mathbf{S}_t)$.

As was mentioned above, we apply the Gaussian copula method to construct the underlying vector process $\mathbf{X}_t^{(\boldsymbol\rho)}$ with transition PDFs of the
form (\ref{urho}) instead of direct coupling of the $S$-space asset
price processes. This is then followed by applying a nonlinear
transformation to obtain the asset prices
$S^{k}_{t}=\mathsf{F}^k(X_{t}^{(\rho_k,k)})$.

To minimize the number of numerical inversions of CDFs, we employ a bridge simulation approach, which is followed by the coupling of the bridge distributions. The application of the Gaussian copula to bridge PDFs from the
$\mathsf{UOU}$ diffusion family leads to a multivariate normal distribution. Indeed, if each CDF $\Phi^k$ (say, the CDF of the $k$th $X^{(\rho)}$-bridge) is a normal CDF $\mathcal{N}\left(\frac{x-\mu_k}{b_k}\right)$, then the multivariate CDF $\mathbf{\Phi}$ given by (\ref{multvarcdf}) with the Gaussian copula in (\ref{normalcop}) is just a multivariate normal distribution function $\mathcal{N}_R\left(\frac{x_1-\mu_1}{b_1},\cdots,\frac{x_n-\mu_n}{b_n}\right)$ with mean vector $(\mu_1,\ldots,\mu_n)^\top$ and covariance matrix $DRD$, where $D=\mathrm{diag}(b_1,\ldots,b_n)$.

\subsection{Path Sampling with a Bridge Normal Copula}
Consider the following back\-ward-in-time bridge sampling algorithm for the exact path-simulation of the
multi-asset price process on a discrete partition $\mathbf{T}$ of the time-interval $[0,T]$. First, we generate the discrete-time random process $(\mathbf{X}_{t_j}^{(\boldsymbol\rho)})_{j=1,2,\ldots,N}$ conditionally on $\mathbf{X}_0^{(\boldsymbol\rho)}=\left(\mathsf{X}^1(S_0^1),\ldots,\mathsf{X}^n(S_0^n) \right),$  and then obtain sample values of $(\mathbf{S}_{t_j})_{j=1,2,\ldots,N}$ by changing variables. Let us denote $X_j^k=X_{t_j}^{(\rho_k,k)}$ and  $S_j^k=S_{t_j}^k$ $k=1,\ldots,n,$ $j=0,1,\ldots,N.$

\begin{enumerate}[Step 1.]
    \item Apply the inverse mapping functions to obtain the initial values:
      \[ X_0^k = \mathsf{X}^k(S_0^k), \quad k=1,2,\ldots,n. \]
    \item Sample the terminal-time value $\mathbf{X}_N=\left(X_N^1,\ldots,X_N^n\right)$ from the copula (\ref{multvarcdf}) by employing numerical inversion of the
    CDFs \[ \Phi^k(X_N^k)=\int_{-\infty}^{X_N^k} p_X^{(\rho_k,k)}(T;X_0^k,x)\,\d x, \quad k=1,2,\ldots,n.\]
      \begin{enumerate}[(i)]
        \item Draw a normal vector $(Z_1,\ldots,Z_n)$ from the
        $n$-variate normal distribution function $\mathcal{N}_R$ with mean vector zero and
        correlation matrix~$R$.
        \item Obtain uniform variates $U_k=\mathcal{N}(Z_k)$,
        $k=1,\ldots,n$.
        \item For each $k=1,\ldots,n$ obtain $X_N^k=(\Phi^k)^{-1}(U_k)$.
      \end{enumerate}
    \item Sample $\mathbf{X}_j=\left(X_j^1,\ldots,X_j^n\right)$ for each $j=N-1,\ldots,1$ by applying the bridge normal copula method as follows.
      \begin{enumerate}[(i)]
        \item Draw a normal vector $(Z_1,\ldots,Z_n)$ from $\mathcal{N}_R$.
        \item For each $k=1,\ldots,n$ set $X^{k}_j=\mu_{kj}+b_{kj}Z_k$, where
        $\mu_{kj}$ and $b^2_{kj}$ are given by (\ref{eqn4ab}) with respective parameters $\lambda=\lambda_k,$ $\kappa=\kappa_k$,
        $\Delta_1=t_j-t_{0},$ and $\Delta_2=t_{j+1}-t_{j}.$
      \end{enumerate}
    \item Map the resulting values of the multivariate discrete-time process $\left(\mathbf{X}_{t_j}^{(\boldsymbol\rho)}\right)$ into the asset prices at each time point:
    \[X^k_j\longrightarrow S^k_j=\mathsf{F}^k(X^k_j),\quad j=1,\ldots,N,\; k=1,2,\ldots,n.\]
\end{enumerate}
As is seen from the algorithm, Step 1 involves the numerical inversion of $n$ distribution functions.
Since all parameters and initial spot prices are fixed,  we can pre-compute and store the values of the inverse CDFs
on a fine grid in $(0,1)$ and then apply a spline interpolation to sample $\mathbf{X}_N$ (see \cite{HORM03}).

Notice that the exact bridge simulation method presented above is faster than any approximation method such as the Euler scheme or the Milstein scheme (e.g., see \cite{Gl04}). First of all, our method has no limitations on time increments which can be very large. Therefore, if a~path needs to be sampled only at a few time moments, no intermediate times are required to guarantee the convergence of sample paths. Second, the Euler approximation method (or any similar one) being applied to an SDE  defined by (\ref{XrhoGen}) requires \emph{frequent} computations of special functions that appear in the drift and diffusion coefficient functions.
The bridge sampling method has only two computationally expensive steps, namely the sampling of terminal asset prices and the computation of mapping functions. As is mentioned above, the respective probability distributions can be tabulated to speed up the sampling. The mapping functions may be tabulated as well.

The approach presented here can be applied to other families of hypergeometric diffusions (see \cite{CM08a,CM08b}).
As is shown in \cite{CM07}, the probability distributions of the squared Bessel process and CIR process, which are the
underlying diffusions for the so-called Bessel and confluent families of $F$-diffusions, respectively, are reduced to randomized
gamma distributions. Such distributions are mixture probability distributions that can be obtained by allowing the shape parameter
of the gamma distribution to be random. To couple $F$-diffusions, one can just couple the gamma distributions using a copula function.
The use of a multivariate gamma distribution is more preferable, but all known examples of such a distribution admit only positive correlations.

\section{Calibration of the Multivariate \textsf{UOU} Model} \label{sect3}

\subsection{Univariate Case} \label{subsect3.1}
It is very important from the practical point of view to develop a reliable and reasonably quick calibration scheme for the \textsf{UOU} diffusion family.
Our objective is to obtain a calibration scheme that provides two levels of calibration: first, an initial full calibration of all parameters of the model
and, second, a much faster recalibration that can be used as soon as new data have arrived. The second calibration scheme may be used throughout the day or even for
longer periods, while the full calibration only need to be executed if markets move considerably.

To estimate a best-fitted parameter set $\xi = \{\lambda,\nu, c, \rho \}$ of the \textsf{UOU} model based on
(observed) market option price data, the least squares method is employed. Suppose that an option with strike $K_i$ and maturity $T_i$
has an observed price $O_i$, while the model produces a price of $C_i=C(K_i, T_i;\xi)$ for the same option, where $i=1,2,\ldots,M$.
The goal of the calibration process is to minimize the least squares error for the $M$ options considered:
\begin{equation}
\label{eq:LSE}
F(\xi) = \sum_{i=1}^M w_i \left|C(K_i, T_i;\xi) - O_i\right|^2 \rightarrow \min_{\xi},
\end{equation}
where $w_i$ is a weight that reflects the relative importance of reproducing the $i$th option price precisely.

The suitable choice of the weight factors $w_i$, $i=1,2,\ldots,M$, is crucial for good calibration results. The confidence in individual data points is determined by the liquidity of the option. The weights can be evaluated from the bid-ask spreads: $w_i=|O_i^\text{ask}-O_i^\text{bid}|^{-1}$. Alternatively, as it was suggested by \cite{CT04}, one may use the Black-Scholes (BS) ``Vegas'' evaluated at the implied volatilities of the market option prices to compute the weights:
$ w_i = \left(\partial C^\text{BS} (\sigma^\text{BS}_i)/\partial \sigma\right)^{-2}$,
where $\partial C^\text{BS}/\partial \sigma$ denotes the derivative of
the BS option pricing formula with respect to the volatility $\sigma$, and $\sigma^\text{BS}_i=\sigma^\text{BS}(O_i,K_i,T_i)$ is the BS implied volatility for the observed market price~$O_i$.

In general, the calibration of a pricing model is an inverse problem, whose solution depends discontinuously on the data.
To achieve uniqueness and stability of the solution, a penalty function is added to the least squares term:
\begin{equation}
\label{eq:LS_reg}
F_{\alpha}(\xi) = \sum_{i=1}^M w_i \left|C(K_i, T_i;\xi) - O_i\right|^2 + \alpha H(\P,\P_0) \rightarrow \min_{\xi},
\end{equation}
where the penalty function $H$ is chosen such that the problem becomes well-posed.

As is examined in \cite{CT04}, the relative entropy method may be applied for solving ill-posed calibration problems. The relative entropy of a probability measure $\P$ on sample space $\Omega$ with respect to some primal measure $\P_0$ is defined as follows:
\begin{equation}
H(\P,\P_0) = E^{\P}\left[\ln \frac{\d \P}{\d \P_0} \right] = \int_{\Omega} \ln \frac{\d \P}{\d \P_0}\,\d \P.
\end{equation}

The regularization parameter $\alpha$ in (\ref{eq:LS_reg}) is used to adjust the trade-off between the accuracy of calibration and the numerical stability of results with respect to input option data. The right choice of $\alpha$ is based on the Morozov discrepancy principle \cite{EH96}, which is described by the following algorithm:
\begin{enumerate}
  \item Compute parameters of $\xi_0$ of the primal measure $\P_0$ by solving the nonlinear least squares problem (\ref{eq:LSE}) in low precision.
  \item Fix $\delta \in (1,1.5)$ and numerically solve equation $F_{\alpha}(\xi_0) = \delta F(\xi_0)$ for the regularization parameter $\alpha$,
   where $F_{\alpha}(\xi_0)$ is defined in (\ref{eq:LS_reg}).
\end{enumerate}

\subsection{Numerical results for the univariate case}\label{subsect3.2}

The data set used consists of $79$ European call option prices with maturities ranging from less than one month up to 1.56 years.
These market prices were obtained from Yahoo for IBM having the spot share value of $101.34$ on July 7th, 2009. For the sake of simplicity,
the risk-free interest is assumed to be constant and equal to $r=0.25\%$, and the dividend rate is set to zero. The calibration routine was
developed using Matlab with the Optimization Toolbox, running on an Intel Core 2 CPU 2.14GHz with 2 GB of main memory.

\begin{figure}
\includegraphics[width=0.49\linewidth]{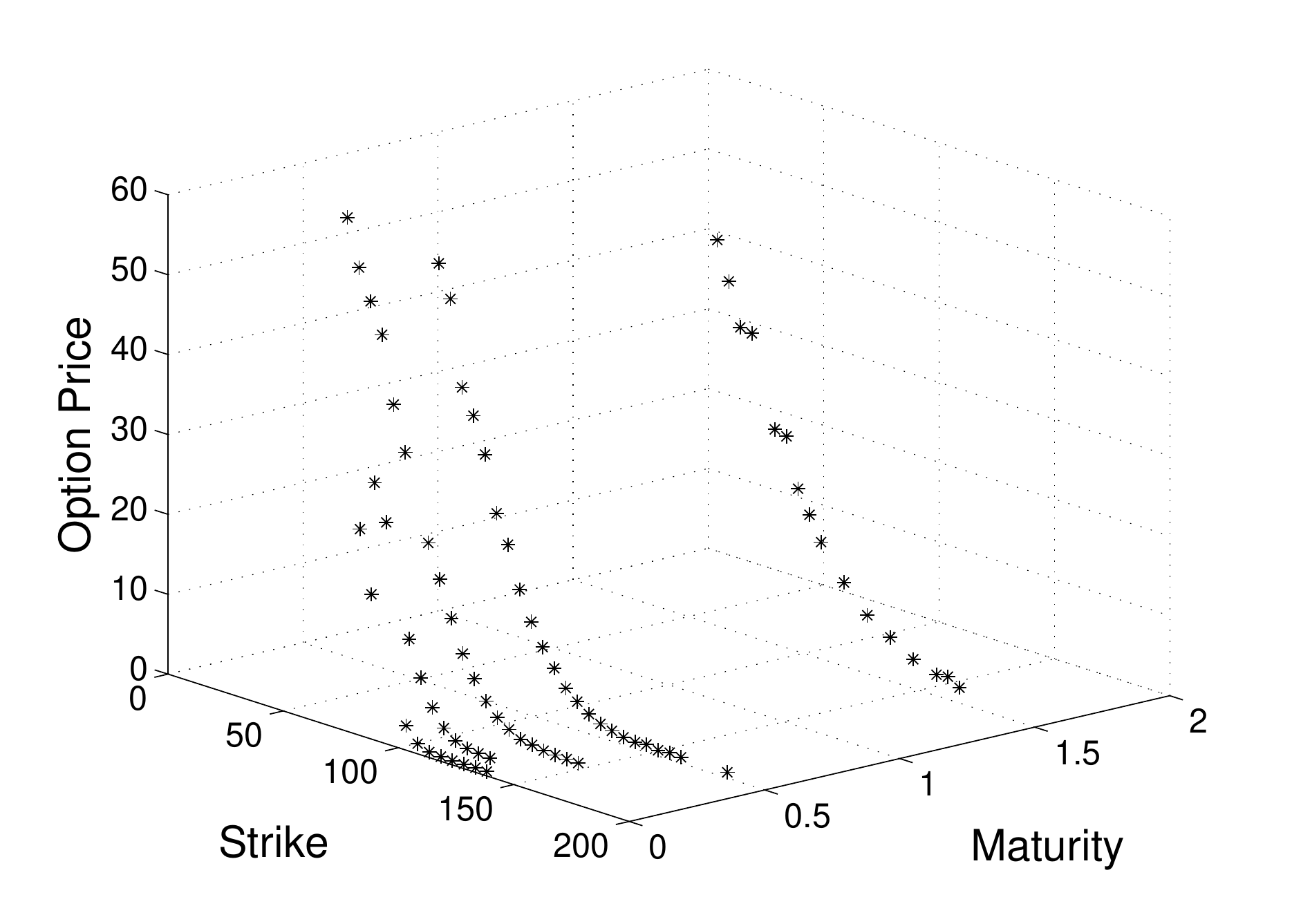}\hfill
\includegraphics[width=0.49\linewidth]{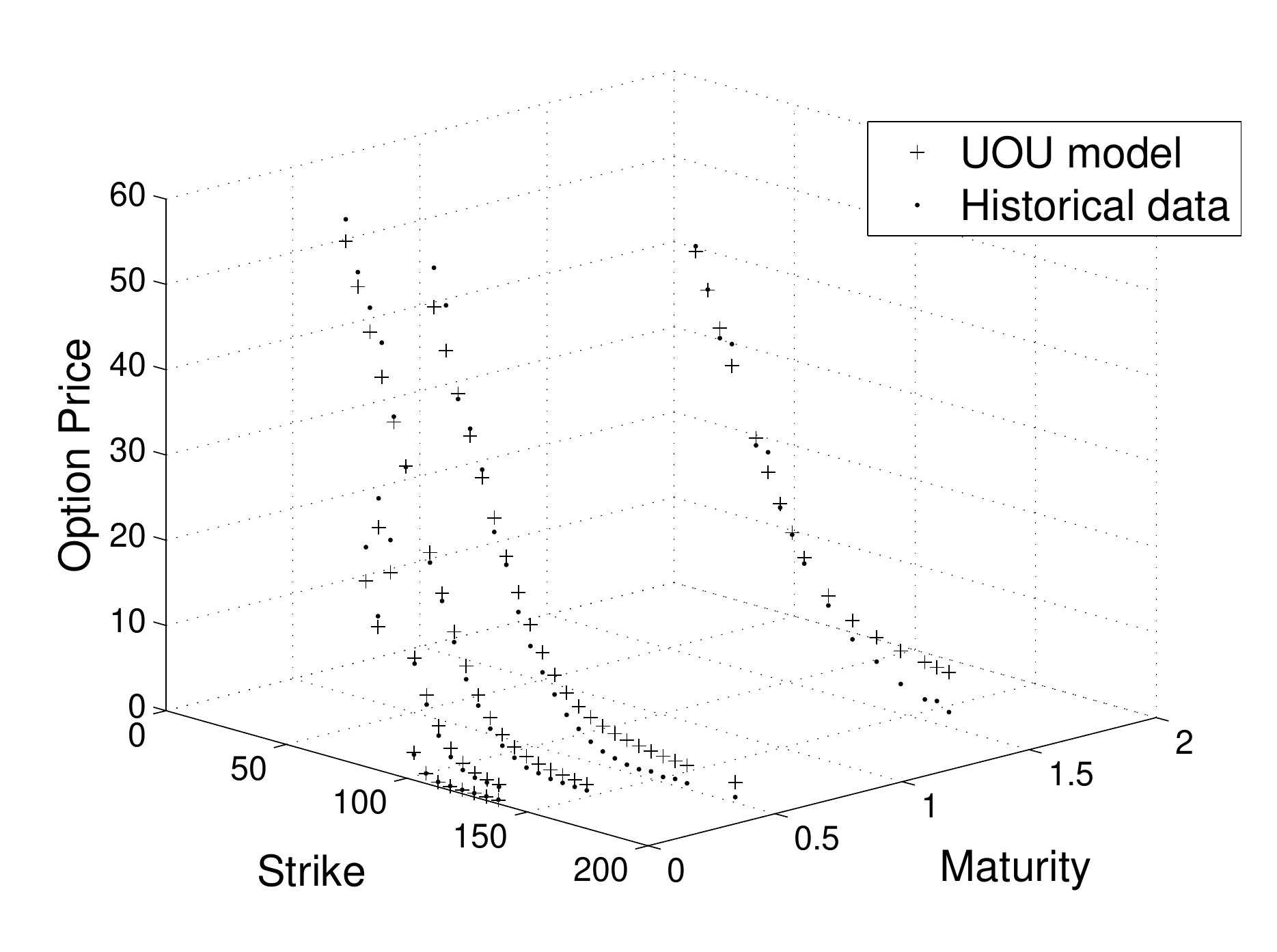}
\vspace*{8pt}
\caption{Market call price surface for IBM, July 7th, 2009 (left plot). Comparison of historical option prices and option prices calculated using
the \textsf{UOU} model with the optimal parameter set (right plot).}
\label{fig:ibm_opt_prices}
\end{figure}

To obtain the set of parameters for the primal probability measure, the \textsf{UOU} model is calibrated to the historical
data from May 7th to July 7th, 2009. Using historical asset prices, $\widehat{S}_{t_j}$, $j = 0,1,\ldots,N$, $0=t_0<t_1<\cdots<t_N,$
and the transition densities, we obtain
the following (single-asset) log-likelihood function for this set of observations:
\begin{equation}
\label{eq:log_MLF}
 \begin{aligned}
L_1(\xi) &= \sum_{j=1}^{N} \ln p_S(t_{j}-t_{j-1};\widehat{S}_{t_{j-1}},\widehat{S}_{t_{j}};\xi)\\
         &= \sum_{j=1}^{N} \ln\left( \frac{\nu}{\sigma(\widehat{S}_{t_{j}};\xi)} p_X^{(\rho)}\left( t_{j}-t_{j-1}; \mathsf{X}(\widehat{S}_{t_{j-1}};\xi), \mathsf{X}(\widehat{S}_{t_{j}};\xi);\xi\right)\right).
 \end{aligned}
\end{equation}
Here, for simplicity, we assume the sequential simulation method. In case of a general sampling method, the log-likelihood function is given by
\begin{equation}
\label{eq:log_MLFgen}
L_1(\xi) = \sum_{j=1}^{N} \ln\left( \frac{\nu}{\sigma(\widehat{S}_{\tilde{t}_{j}};\xi)} f_j\left(\mathsf{X}(\widehat{S}_{\tilde{t}_{j}};\xi);\xi\right)\right),
\end{equation}
where $f_j$ is defined analogously to $f_j^k$ in Subsection~\ref{subsect2.3} as if there was only one asset.

In practice, the implementation of the calibration procedure is started with some initial values of parameters.
The upper and lower bounds for the parameters  should also be provided. Based on the empirical analysis, such bounds are obtained and are provided
in Table~\ref{tbl:par_set}.

\begin{table}
  \caption{Initial values and bounds for the parameters of the \textsf{UOU} model.}
  \label{tbl:par_set}
  \begin{center}
   \begin{tabular}{lrrrr}
\hline
{Parameter} &  { $\rho$} &    { $\upsilon$} &    { $c$} &    { $\kappa$} \\
\hline
{ Lower bound} &      0.001 &       0.005 &          45 &        0.5 \\
{ Upper bound} &          0.5 &          2 &        250 &         10 \\
\hline
{ Initial value} &          0.04 &          0.34&        102.59 &         1\\
\hline
\end{tabular}
  \end{center}
\end{table}

The first step of the calibration procedure takes approximately $200$ seconds to fit the model to $63$ historical asset prices.
The optimal values, that maximize the log-likelihood function (\ref{eq:log_MLF}), are $\rho = 0.0357$, $\upsilon = 0.0531$,
$c =118.2404$, $\kappa = 0.5951$. This set of parameters defines the primal probability measure $\P_0$. The estimation of the regularization parameter
$\alpha$ is based on the algorithm described above. The calculated value of $\alpha$ is $0.266$.

The final step of the calibration process is the minimization of the nonlinear least squares function regularized by
the relative entropy as is given in (\ref{eq:LS_reg}). The computation algorithm utilizes the Matlab function \textsf{lsqnonlin}
with the exit tolerance set to $10^{-6}$. This function employs the Levenberg-Marquardt least-squares algorithm for estimating
optimal parameters. The starting values and the limits for the parameters remain the same as given in Table~\ref{tbl:par_set}.
The computational algorithm takes approximately 400 seconds to fit the model to 79 option prices. The best-fitted parameters of the model are
$\rho = 0.0203$, $\upsilon = 0.0013$, $c =102.1384$, $\kappa = 0.6579$. The objective function $F_{\alpha}$ attains its minimum value of 1.58.

Notice that the discrepancy between the computed option prices and observed option prices may originate from different sources.
First, the market data may contain errors or misleading information. For example, the values of illiquid options might be mispriced,
or simple input errors may occur. Second, the calibration procedure estimates model parameters of an arbitrage-free model, while the market
prices are not necessarily arbitrage-free. Hence, there is an inherent mismatch between the model prices and the market data. Notice that the use
of time-dependent parameters may decrease the level and number of errors and make the calibration procedure maturity-wise. Another possible solution
to improve the accuracy is to employ the calibration separately for out-of-the-money, at-the-money and in-the-money options.

\subsection{Multivariate Case}\label{subsect3.3}

Let us consider the multi-asset price processes $(\mathbf{S}_t)_{t\ge0}$ modeled as described in Section~\ref{sect2}, i.e. $n$ \textsf{UOU} diffusions are coupled via the Gaussian copula function.

The calibration procedure can be split into two stages: (1) estimation of the parameters of the marginal (single-asset price) processes; (2) estimation of the correlation matrix $R$ of the Gaussian copula. Such a calibration algorithm admits multiple variations. First, one may use the maximum likelihood estimation (MLE) to fit the marginal models to historical asset prices. Second, one may use the least-square method to fit the marginal models to historical derivative prices (say European options). For both approaches, the correlation matrix
is then estimated by the MLE using historical asset prices. Alternatively, one may use only observed asset prices to estimate all parameter of the multivariate model simultaneously without splitting the calibration process. Notice also that the multivariate path distribution depends on the simulation method used. By using the sequential sampling or some version of the bridge sampling, one may obtain different models and, hence, obtain slightly different estimates of the model parameters.

Let $\left\{\mathbf{S}_j\equiv( \widehat{S}_{t_j}^1,\ldots,\widehat{S}_{t_j}^n)\right\}_{j=0}^N$ be the $n \times (N+1)$ matrix containing $N+1$
independent historical prices for each of the $n$ financial assets observed on a~set of time points $T=\{t_0,t_1,\ldots,t_N\}$. Let
$({\boldsymbol\xi},R) =(\xi_1,\ldots,\xi_n,R)$ denote the set of parameters to be estimated. The historical observations in $X^{(\rho)}$-space are
obtained by applying the inverse map: $\widehat{X}^{k}_{t_j} = \mathsf{X}^k(\widehat{S}_{t_j}^{k};\xi_k)$. Suppose the joint path PDFs of
the $n$-dimensional processes $(\mathbf{X}_t^{(\boldsymbol\rho)})$ and $(\mathbf{S}_t)$ is constructed with the Gaussian copula as given by (\ref{pathDF}) and (\ref{jointpathSPDF}), respectively. The ($n$-asset) log-likelihood function is
\begin{equation}
\label{eq:MLE_1}
\begin{split}
L_n({\boldsymbol\xi},R) &= \ln\mathbb{f}^s(\mathbf{S}_1,\ldots,\mathbf{S}_N) = \sum_{j=1}^N \ln \mathbf{f}^s_j(\mathbf{S}_j)\\ &=\sum_{j=1}^N \ln
\phi_R(\mathcal{N}^{-1}(\Phi^1_j(\widehat{X}_{\tilde{t}_j}^1;\xi_1)),\ldots,\mathcal{N}^{-1}(\Phi^n_j(\widehat{X}_{\tilde{t}_j}^n;\xi_n)))\\
 &\phantom{=} + \sum_{k=1}^n \sum_{j=1}^N \ln \left(\frac{\nu_k}{\sigma^k(\widehat{S}_{\tilde{t}_j}^k)} f_j^k(\widehat{X}_{\tilde{t}_j}^k;\xi_k)\right)\equiv L_n^{\text{corr}}(R|{\boldsymbol\xi})+\sum_{k=1}^n L_1(\xi_k),
\end{split}
\end{equation}
where $\phi_R$ denotes the joint PDF of the $n$-variate normal distribution with mean vector zero, unit variances, and correlation matrix $R$; $L_1$ is the single-asset log-likelihood function given by (\ref{eq:log_MLF}) or (\ref{eq:log_MLFgen}), and $L_n^{\text{corr}}$ denotes the log-likelihood function for the copula function.
Recall that the expression in (\ref{eq:MLE_1}) is independent of the simulation method used. For the sequential and bridge methods, we
provide below specific expressions of the log-likelihood function.


As is suggested by the structure of the log-likelihood function in (\ref{eq:MLE_1}), the calibration process can be split into two steps. First, the sets $\xi_k=\left\{\lambda_k,\nu_k,c_k,\rho_k\right\},\,
k = 1,2,\ldots,n$ of parameters of the marginal distributions are estimated by employing the maximum likelihood estimation:
\begin{equation}
\label{eq:MLE_margins}
\widehat{\xi}_k = \mathrm{arg} \max_{\xi_k} L_1(\xi_k)=\mathrm{arg} \max_{\xi_k}\sum_{j=1}^{N} \ln \left(\frac{\nu_k}{\sigma^k(\widehat{S}_{\tilde{t}_j}^k)} f_j^k(\widehat{X}_{\tilde{t}_j}^k;\xi_k)\right), \ \ k = 1,\ldots,n.
\end{equation}

As is seen, the parameters of the marginal distributions are estimated based on historical data. An alternative
approach to computing the parameters is to fit asset price distributions to observed option prices as described in Subsection~\ref{subsect3.1}.

The last step is the estimation of the correlation matrix $R$ for the given optimal univariate model parameters
$\mathbf{\widehat{\boldsymbol\xi}} \equiv \{\widehat{\xi}_1,\ldots,\widehat{\xi}_n \}$, $\widehat\xi_k\equiv\left\{\hat\lambda_k,\hat\nu_k,\hat c_k,\hat\rho_k\right\}$, estimated during the previous step.

First, we consider the sequential calibration method with the following log-likelihood function
\begin{equation} \label{eq:MLE_corr_seqn}
 L_n^\text{corr}(R|\widehat{\boldsymbol{\xi}}) =  \sum_{j=1}^{N} \ln
\phi_R(\mathcal{N}^{-1}(\Phi^1_j(\widehat{X}_{{t}_j}^1;\widehat\xi_1)),\ldots,\mathcal{N}^{-1}(\Phi^n_j(\widehat{X}_{{t}_j}^n;\widehat\xi_n))).
\end{equation}

\paragraph{\textbf{Sequential Calibration.}}
For the sequential path generation method, the algorithm is as follows.
\begin{enumerate}[(i)]
    \item Map all the observations into $X^{(\rho)}$-space using the respective inverse maps:
            \[
                 \mathbf{X}_j \equiv  (\widehat{X}_j^1,\ldots,\widehat{X}_j^n) =  ((\mathsf{X}^1(\widehat{S}_{t_j}^1;\widehat{\xi}_1),\ldots,\mathsf{X}^n(\widehat{S}_{t_j}^n;\widehat{\xi}_n)),\ \ j=0,\ldots,N.
            \]
   \item Compute vectors $\mathbf{u}_j \equiv  (u_j^1,\ldots,u_j^n) \in [0,1]^n,$ $j=1,\ldots,N,$ by evaluating the integrals:
                \[u^k_j = \int_{-\infty}^{\widehat{X}_j^k} p_X^{(\hat\rho_k,k)}(t_i-t_{j-1};\widehat{X}_{j-1}^k,x;\widehat\xi_k) \,\d x,\ \ k=1,\ldots,n\]
   \item Maximize the log-likelihood function with respect to $R$:
    \[
    L_n^\text{corr}(R|\widehat{\boldsymbol{\xi}}) = \sum_{j = 1}^{N} \ln \phi_R\left(\mathcal{N}^{-1}(u_j^1),\ldots,\mathcal{N}^{-1}(u_j^n)\right)\to\max_R.
    \]
\end{enumerate}

The estimation of the log-likelihood function for the sequential calibration involves numerous estimations of the CDF for the \textsf{UOU} model. Since there is no simple-form solution for the CDF, the numerical integration of the probability density function should be performed regularly. By applying the bridge approach to the construction of the multivariate path distribution function, the number of integrals to be computed numerically on step (ii) reduces from $n \times N$ to $n$. This is due to the fact that for the bridge approach, the CDF $\Phi^k_j$, $j=2,
 \ldots,N$, is Gaussian. Hence, for the bridge path generation method, the log-likelihood function can be simplified as follows:
  \begin{equation} \label{eq:MLE_corr_bridge}
  \begin{split}
 L_n^\text{corr}(R|\widehat{\boldsymbol{\xi}}) &=  \sum_{j=2}^{N} \ln \phi_R\left(\dfrac{\widehat{X}_{\tilde{t}_j}^1 - \mu_{kj}}{b_{kj}},\ldots,\dfrac{\widehat{X}_{\tilde{t}_j}^n - \mu_{kj}}{b_{kj}}\right) \\
   &\phantom{=} + \ln \phi_R(\mathcal{N}^{-1}(\Phi_1^1(\widehat{X}_{\tilde{t}_1}^1;\widehat{\xi}_1)),\ldots,\mathcal{N}^{-1}(\Phi_1^n(\widehat{X}_{\tilde{t}_1}^n;\widehat{\xi}_n))),
 \end{split}
\end{equation}
where mean $\mu_{kj}$ and variance $b^2_{kj}$ computed by formulae in (\ref{eqn4ab}); and $\tilde{t}_1=T$.


\paragraph{\textbf{Bridge Calibration.}} The following algorithm can be applied for the backward-in-time bridge path generation method.
\begin{enumerate}[(i)]
 \item Map the observations into $X^{(\rho)}$-space $\mathbf{S}_j \rightarrow \mathbf{X}_j = \mathbf{F}^{-1}(\mathbf{S}_j;\widehat{\boldsymbol\xi}), j=1,\ldots,N$, as is described in part~(i) of the sequential algorithm.
  \item Compute $\mathbf{u}_{N}\equiv (u_N^1,\ldots,u_N^n)$, the values of normal CDFs corresponding to the terminal point of a path:
  \begin{equation}
  \begin{split}
  	u_N^k =& \int_{-\infty}^{\widehat{X}^k_N} p_X^{(\hat\rho_k,k)}(t_N - t_{N-1};\widehat{X}_{N-1}^k,x;\widehat\xi_k)\,\d x, \ \ k=1,\ldots,n.\\
  \end{split}
  \end{equation}
  \item For each $k=1,\ldots,n$ and $j=1,\ldots ,N-1$ calculate
        $\mu_{kj}$ and $b_{kj}$ by using (\ref{eqn4ab}) with respective parameters $\lambda=\lambda_k,$ $\kappa=\kappa_k$,
        $\Delta_1=t_j-t_{0},$ and $\Delta_2=t_{j+1}-t_{j}.$ Then, set
        \[x^k_j = \frac{\widehat{X}_j^k-\mu_{kj}}{b_{kj}}.\]
 \item Maximize the log-likelihood function with respect to $R$:
    \[
    L_n^\text{corr}(R|\widehat{\boldsymbol{\xi}}) = \sum_{j = 1}^{N-1} \ln \phi_R\left(x^1_j,\ldots,x^n_j\right) + \ln \phi_R \left(\mathcal{N}^{-1}(u_N^1),\ldots,\mathcal{N}^{-1}(u_N^n)\right)\to\max_R.
    \]
\end{enumerate}

\begin{table}[ht]
  \caption{Optimal parameters estimated for IBM, Microsoft, Pepsi and Wallmart}
  \label{tbl:cali_hist_res_4dim}
  \begin{center}
  \small{
\begin{tabular}{l|rrrrr}
            & IBM & Microsoft & Pepsi & Wallmart\\
\hline
       $\hat\rho$  &     0.0496   &    0.2173 & 0.0865 & 0.0493   \\
       $\hat\upsilon$   &     0.0887   &    0.0365 & 0.1149 & 0.0886  \\
       $\hat c$   &   103.9904   &    21.1638& 31.671 & 52.3842  \\
     $\hat\kappa$  &     0.9670   &    0.874  & 0.910  & 0.9874
\end{tabular}}
  \end{center}
\end{table}

\subsection{Numerical results for the multivariate case}\label{subsect3.4}

For this numerical experiment the daily observations of four American companies, namely, IBM, Microsoft, Pepsi, and Wallmart,
have been collected from \textsc{Yahoo!}$^{\text{TM}}$. The examined period is April 7th, 2009, to July 7th, 2009, and it consists of 63 time points.
In the first stage of the calibration, the optimal sets of parameters of the marginal distributions are estimated by solving Eq.~(\ref{eq:MLE_margins}),
and they are provided in Table~\ref{tbl:cali_hist_res_4dim}.

Two approaches are then used for the evaluation of the optimal correlation matrix $R$.
In the first approach, the correlation matrix is obtained by the pairwise calculation of the correlation coefficients.
There are $\binom{4}{2}$ correlation coefficients for $4$-asset price processes to be estimated. In other words, instead of solving a $4$-dimensional least-square problem with the log-likelihood function $L_n^\text{corr}$ given by (\ref{eq:MLE_corr_seqn}) or (\ref{eq:MLE_corr_bridge}), we independently solve $\binom{4}{2}$ $2$-dimensional problems with a 2-by-2 correlation matrix of the form $\left(\begin{array}{cc} 1 & \theta\\ \theta & 1 \end{array} \right)$ and then compose another 4-by-4 correlation matrix using the estimations of $\theta$.

 However, the resulting matrix
may violate the positive-definite property. To overcome this problem, a method suggested by \cite{JR99} of finding the closest
correlation matrix by the spectral decomposition is applied. The resulting matrices $\widehat{R}$ are shown in Table \ref{tbl:cali_res_4dim}. The
computation time is $32.6$ seconds for the bridge simulation and $47.9$ seconds for the sequential simulation.

\begin{table}[ht]
  \caption{Correlation matrices obtained by using the bridge path simulation (left table) and the sequential
path simulation (right table). The pairwise computation of the correlation coefficients is employed.}
  \label{tbl:cali_res_4dim}
\begin{minipage}[b]{0.475\linewidth}
\centering
  $$
  \small{
\left(\begin{array}{rrrr}
     1 &      0.297 &      0.151 &      0.337 \\
     0.297 &      1 &      0.089 &     -0.045 \\
     0.151 &      0.089 &      1 &     -0.080 \\
     0.337 &     -0.045 &     -0.080 &      1 \\
\end{array}\right)};
$$
\end{minipage}\hfill
\begin{minipage}[b]{0.475\linewidth}
\centering
  $$
  \small{
\left(\begin{array}{rrrr}
      1 &      0.278 &      0.243 &      0.336 \\
     0.278 &      1 &      0.184 &     -0.050 \\
     0.243 &      0.184 &      1 &     -0.051 \\
     0.336 &     -0.050 &     -0.051 &      1 \\
\end{array}\right)}.
$$
\end{minipage}

\end{table}

In the second method, the correlation matrix as a whole is estimated. The computation of an optimal correlation matrix is
performed in Matlab using the function \textsf{fmincon}, which allows us to find a minimum of a multivariate function with
non-linear constraints. By adding nonlinear constraints, the algorithm works in the class of semi-positive matrices, which
is absolutely necessary for the correct formulation of the correlation matrix. However, the candidate matrix, which minimizes
the objective function in (\ref{eq:MLE_1}), may not have ones on the principal diagonal. To obtain a correct correlation matrix
that is closest to the given one, the spectral decomposition method is applied again. The results are shown in Table~\ref{tbl:cali_res_4dim_2}
and Table~\ref{tbl:cali_res_4dim_3}.

\begin{table}[ht]
 \caption{The candidate semi-positive matrix that minimizes the objective function $L_n^\text{corr}$ in (\ref{eq:MLE_corr_seqn}) (left table) obtained by using the
sequential simulation method and the closest correlation matrix obtained by using the spectral decomposition method (right table).}
  \label{tbl:cali_res_4dim_2}
\begin{minipage}[b]{0.45\linewidth}
\centering
  $$
  \small{
\left(\begin{array}{rrrr}
      1.000 &      0.277 &      0.169 &      0.335 \\
     0.277 &      0.993 &      0.127 &     -0.049 \\
     0.169 &      0.127 &      0.480 &     -0.024 \\
     0.335 &     -0.049 &     -0.024 &      0.993 \\
\end{array}\right)}
$$
\end{minipage}
\hfill$\underset{\text{decomp.}}{\overset{\text{spect.}}{\longrightarrow}}$\hfill
\begin{minipage}[b]{0.45\linewidth}
\centering
  $$
  \small{
\left(\begin{array}{rrrr}
     1 &      0.278 &      0.243 &      0.336 \\
     0.278 &      1 &      0.183 &     -0.049 \\
     0.243 &      0.183 &      1 &     -0.035 \\
     0.336 &     -0.049 &     -0.035 &      1 \\
\end{array}\right)}
$$
\end{minipage}
 \end{table}

\begin{table}[ht]
  \caption{The candidate semi-positive matrix that minimizes the objective function $L_n^\text{corr}$ in (\ref{eq:MLE_corr_bridge}) (left table) obtained by using the bridge simulation method  and the closest correlation matrix obtained by using the spectral decomposition method (right table).}
  \label{tbl:cali_res_4dim_3}
\begin{minipage}[b]{0.45\linewidth}
\centering
  $$
  \small{
\left(\begin{array}{rrrr}
     1.000 &      0.290 &      0.145 &      0.335 \\
     0.290 &      0.973 &      0.084 &     -0.043 \\
     0.145 &      0.084 &      0.966 &     -0.076 \\
     0.335 &     -0.043 &     -0.076 &      0.993 \\
\end{array}\right)}
$$
\end{minipage} \hfill$\underset{\text{decomp.}}{\overset{\text{spect.}}{\longrightarrow}}$\hfill
\begin{minipage}[b]{0.45\linewidth}
\centering
  $$
  \small{
\left(\begin{array}{rrrr}
     1 &      0.293 &      0.148 &      0.336 \\
     0.293 &      1 &      0.086 &     -0.043 \\
     0.148 &      0.086 &      1 &     -0.078 \\
     0.336 &     -0.043 &     -0.078 &      1 \\
\end{array}\right)}
$$
\end{minipage}
\end{table}


\section{Pricing Path-Dependent Options} \label{sect4}
\nopagebreak
\subsection{Path-Dependent Multi-asset Options} \label{subsect4.1}
\nopagebreak
Suppose the price process $(\mathbf{S}_t)$ is modelled at a discrete
set of time points (written in an increasing order)
$\mathbf{T}=\{0=t_0,t_1,t_2,\ldots,t_N=T\}$. Hence, we construct a
multivariate discrete-time $n$-dimensional price path $(\mathbf{S}_i)_{i=1,2,\ldots,N},$
where $\mathbf{S}_j=(S_j^{1},S_j^{2},\ldots,S_j^{n})$.

Let us consider two discrete-time monitored  path-dependent securities, namely a
Bermudan option and an Asian option. For an Asian-style option the
payoff function $\Lambda_N(\mathbf{S}_1,\ldots,\mathbf{S}_N)$ is
assumed to be a function of averages
$A^{k}=\mathcal{A}\left(S_1^{k},\ldots,S_N^{k}\right)$ of the asset
prices, where, for example in the case of the arithmetic
time-averaging, ${A}^{k}=\frac{1}{N}\sum_{i=1}^N S_i^{k}$. The Asian
basket option with the payoff \begin{equation} \label{AsBskCall}
\Lambda_N=\left( \max\limits_{k=1}^n A^{k}-K\right)^+,\end{equation}
considered in Section \ref{subsect4.2}, is an example of an
arithmetic average option.

The arbitrage-free value $V=V(\mathbf{S}_0,\mathcal{T})$ of a discrete-time monitored
path-dependent option (without early exercise opportunities) can be
represented as a mathematical expectation under a risk-neutral probability measure $\Q$:
\begin{equation} \label{mvexpect}
 V = e^{-r T}\E^\Q[\Lambda_N(\mathbf{S}_1,\ldots,\mathbf{S}_N)] = e^{-r T}\E^\Q[
 \Lambda_N(\mathbf{F}(\mathbf{X}_1),\ldots,\mathbf{F}(\mathbf{X}_N))]\,.
\end{equation}
To find the option value one may use the Monte Carlo method.

The fair value of a Bermudan option cannot be represented in
the form (\ref{mvexpect}), but is a solution of some dynamic
programming problem (see (\ref{dynprog}) below). Let
$\Lambda(t,\mathbf{S})$ denote the payoff function for exercise at
time $t$ in state $\mathbf{S}=\left(S^{1},\ldots,S^{n}\right)$. One
may consider the two following examples of the time-homogeneous
payoff function~$\Lambda$:
\begin{itemize}
    \item a max call option with the payoff
$\Lambda(\mathbf{S})=\left(\max(S^{1},\ldots,S^{n})-K\right)^+$;
    \item a geometric average call with the payoff
$\Lambda(\mathbf{S})=\left(\big(\prod\limits_{k=1}^n
S^{k}\big)^{\frac{1}{n}}-K\right)^+$.
\end{itemize}

Let $V(t,\mathbf{S})$ denote the value of the option at time $t$
given $\mathbf{S}_t=\mathbf{S}$, assuming the option has not
previously been exercised. We are interested in the present value
$V(t_0,\mathbf{S}_0)$. This value is determined through the
backward-in-time recursion:
\begin{equation} \label{dynprog}
  \begin{array}{rcl}
    V(t_N,\mathbf{S}) &=& \Lambda(t_N,\mathbf{S}) \\
    V(t_i,\mathbf{S}) &=& \max\left\{ \Lambda(t_i,\mathbf{S}),
    \E^\Q\left[e^{-r (t_{i+1}-t_i)} V(t_{i+1},\mathbf{S}_{i+1}) \mid \mathbf{S}_i=\mathbf{S} \right]
    \right\}\,,
  \end{array}
\end{equation}
where $i=N-1,\ldots,0$. The risk-neutral expectation value function
in the latter expression is called the continuation value function
in state $\mathbf{S}$ at time $t_{i-1}$.


In our computational tests, the univariate models
correspond to the model parameter values used to plot the local
volatility function with the thick line in Figure~\ref{Fig1}. So,
for the $\mathsf{UOU}$-family we simply set $\rho_k\equiv 0.02$,
$\kappa_k=1$, $\upsilon_k=0.5$ and $c_k=100$, for
all $k$. The interest rate is $r=0.05$.

\subsection{Pricing Asian Basket Options} \label{subsect4.2}
In this example we use the Monte Carlo method for pricing the
arithmetic Asian basket call option with payoff function
(\ref{AsBskCall}) under the multivariate \textsf{UOU} model. The Monte Carlo simulations were run on the
SHARCNET network (\url{http://www.sharcnet.ca}). Mostly we use \textsf{Gulper}---a 42-CPU cluster. The code was
written in \textsf{FORTRAN-90} using the \textsf{MPI} library.

First, we present numerical results for a bivariate ($n=2$) \textsf{UOU}
model. The correlation matrix used in the normal
(Gaussian) copula has the usual form of $R=\left(\begin{array}{cc} 1& \theta \\
\theta & 1 \end{array}\right)$. Table \ref{table1} provides the results of our numerical tests for the two-asset model. We used\
the following values of parameters: number of observations $N=100$,
terminal time $T=1.0$, number of scenarios $M=10\,000\,000$, and spot
value $S_0=100$.

\begin{table}
  \centering
  \caption{\label{table1} Pricing the Asian basket call option under the bivariate \textsf{UOU} model
  for different values of the correlation coefficient $\theta$ using the Monte Carlo method.
  A~sample standard error is given after $\pm$ sign.}
  \begin{tabular}{rr@{\,$\pm$\,}l@{\quad}r@{\,$\pm$\,}l@{\quad}r@{\,$\pm$\,}l}
     \hline
      $K$ & \multicolumn{2}{c}{$\theta=-0.75$} & \multicolumn{2}{c}{$\theta=0$} & \multicolumn{2}{c}{$\theta=0.75$} \\
     \hline
      90  & 31.538 & 0.008 & 27.942 & 0.008 & 22.444 & 0.009  \\
      100 & 22.786 & 0.007 & 20.409 & 0.008 & 16.168 & 0.007\\
      110 & 15.614 & 0.007 & 14.348 & 0.007 & 11.321 & 0.006 \\
     \hline
  \end{tabular}
\end{table}

The next example confirms that the computational complexity of
the bridge copula method scales linearly with the number of
assets. For all tests we use the crude Monte Carlo method on a single CPU. Clearly, the computational cost can be
significantly reduced by using variance reduction methods and low-discrepancy point sets along with the parallel
computing (see \cite{CM08a}). Another method of speeding up the computer code is to tabulate all distribution,
generating and mapping functions that depend on the parabolic cylinder functions and other special functions.

Table \ref{table2} provides the results of our numerical tests for increasing numbers of assets. We use
the same values of parameters as those of the previous test but reduce the number of scenarios to $M=1\,000\,000$.
A 10-by-10 correlation matrix $R$ given in Figure~\ref{corrR} is generated using the random Gram matrix method from \cite{Ho91}.
For values of $n<10$ we let $R$ be the $n\times n$-submatrix in the upper-left corner of the matrix given in Figure~\ref{corrR}.
Clearly, such a submatrix is again a correlation matrix of lower dimensionality.

\begin{figure}[h]
  \[
  {\small
    \left( \begin{array}{r@{\,\,}r@{\,\,}r@{\,\,}r@{\,\,}r@{\,\,}r@{\,\,}r@{\,\,}r@{\,\,}r@{\,\,}r}
  1.000&  0.550& -0.123&  0.148& -0.239&  0.238&  0.239&  0.096&  0.289& -0.325\\
  0.550&  1.000& -0.223&  0.188& -0.109&  0.396& -0.132& -0.118&  0.237& -0.390\\
 -0.123& -0.223&  1.000&  0.198&  0.135&  0.242&  0.013& -0.731&  0.244& -0.134\\
  0.148&  0.188&  0.198&  1.000& -0.519&  0.314& -0.191& -0.332&  0.469&  0.194\\
 -0.239& -0.109&  0.135& -0.519&  1.000& -0.279&  0.292& -0.044& -0.077& -0.443\\
  0.238&  0.396&  0.242&  0.314& -0.279&  1.000&  0.023&  0.050&  0.049& -0.062\\
  0.239& -0.132&  0.013& -0.191&  0.292&  0.023&  1.000&  0.423& -0.306& -0.292\\
  0.096& -0.118& -0.731& -0.332& -0.044&  0.050&  0.423&  1.000& -0.417&  0.152\\
  0.289&  0.237&  0.244&  0.469& -0.077&  0.049& -0.306& -0.417&  1.000& -0.534\\
 -0.325& -0.390& -0.134&  0.194& -0.443& -0.062& -0.292&  0.152& -0.534&  1.000
\end{array} \right) } \]
 \caption{The 10-by-10 randomly generated correlation matrix $R$.}\label{corrR}
\end{figure}

The idea of \cite{Ho91} is to generate independent pseudo-random vectors $\mathbf{u}_1,\ldots,\mathbf{u}_n$ distributed uniformly on the $n$-dimensional unit sphere and then to use the Gram matrix $R=U^\top U$, where $U\equiv(\mathbf{u}_1|\cdots|\mathbf{u}_n)$ has $\mathbf{u}_k$ as $k$th column and $U^\top$ is the transpose of $U$. To create $\mathbf{u}_k$ in $\R^n$ we use a vector of independent standard normals $\mathbf{z}_k\sim\mathcal{N}(0,I)$ normalized by the Euclidian norm: $\mathbf{u}_k=\mathbf{z}_k / \|\mathbf{z}_k\|.$

\begin{table}
  \centering
  \caption{\label{table2} Pricing the Asian basket call option under the $n$-variate \textsf{UOU} model using the Monte Carlo method.
  A~sample standard error is given after $\pm$ sign.}
  \begin{tabular}{rr@{\,$\pm$\,}l@{\quad}r@{\,$\pm$\,}l@{\quad}r@{\,$\pm$\,}l@{\quad}r@{\,$\pm$\,}l}
     \hline
      $K$ & \multicolumn{2}{c}{$n=2$} & \multicolumn{2}{c}{$n=3$} & \multicolumn{2}{c}{$n=5$} &
      \multicolumn{2}{c}{$n=10$} \\
     \hline
      90  & 24.292 & 0.026 & 34.420 & 0.027 & 45.504 & 0.028 & 59.785 & 0.028 \\
      100 & 17.627 & 0.024 & 25.974 & 0.026 & 36.195 & 0.027 & 50.274 & 0.028 \\
      110 & 12.403 & 0.021 & 18.786 & 0.024 & 27.507 & 0.026 & 40.800 & 0.028 \\
     \hline
      Time & \multicolumn{2}{c}{3\,745 sec} & \multicolumn{2}{c}{5\,340 sec} & \multicolumn{2}{c}{8\,389 sec} & \multicolumn{2}{c}{15\,307 sec}\\
     \hline
  \end{tabular}
\end{table}

\subsection{Pricing Bermudan Options} \label{subsect4.3}
Regression-based methods are broadly applied to pricing multivariate
American options (see \cite{Gl04} and references therein). The main
idea consists in the use of regression to estimate continuation
values from simulated paths. Each continuation value is approximated
by a linear combination of some basis functions. The coefficients of
such a representation are estimated by a regression method
(typically least-squares).

Consider an expression for the continuation value of the form:
\begin{equation} \label{contvalreg}
 \E^\Q\left[ e^{-r (t_{i+1}-t_i)} V(t_{i+1},\mathbf{S}_{i+1}) \mid \mathbf{S}_i=\mathbf{S}
 \right] = \sum\limits_{q=1}^L \beta_{iq}\psi_q(\mathbf{S})=\beta_i^\top\psi(\mathbf{S}),
\end{equation}
for some basis functions $\psi=(\psi_1,\ldots,\psi_L)^\top$,
$\psi_q:\R^n \to \R$ and constants $\beta_{iq}$,
$q=1,2,\ldots,L$, $i=0,1,\ldots,N-1$. An approximate value $\widehat{V}_0$ for the American (i.e. Bermudan)
option price can be calculated by the following regression
algorithm.

\paragraph{\textbf{Regression-Based Pricing Algorithm.}}
\nopagebreak
\begin{enumerate}
    \item[(i)] Simulate $M$ independent asset price paths
$(\mathbf{S}_{1j},\ldots,\mathbf{S}_{Nj}),$ $j=1,\ldots,M.$
    \item[(ii)] At terminal nodes, set
    $\widehat{V}_{Nj}=\Lambda(t_N,\mathbf{S}_{Nj}),$ $j=1,\ldots,M$.
    \item[(iii)] Apply backward-in-time induction for $i=N-1,\ldots,1$:
     \begin{itemize}
        \item Given estimated values $\widehat{V}_{i+1,j},$
        $j=1,\ldots,M,$ the least-squared estimate of
        $\beta_{i}=(\beta_{i1},\ldots,\beta_{iR})^\top$ is given by
        $\widehat\beta_i=\widehat{B}^{-1}_{\psi,i} \widehat{B}_{\psi V,i}\,,$ where
        $\widehat{B}_{\psi,i}$ is the $L\times L$ matrix with $(r,q)$-entry
        $\frac{1}{M}\sum\limits_{j=1}^M
        \psi_r(\mathbf{S}_{ij})\psi_q(\mathbf{S}_{ij})$ and $\widehat{B}_{\psi
        V,i}$ is the $L$-vector with $q$th entry
        $\frac{e^{-r (t_{i+1}-t_i)}}{M}\sum\limits_{j=1}^M
        \psi_q(\mathbf{S}_{ij})\widehat{V}_{i+1,j}.$
        \item Set
        $\widehat{V}_{ij}=\max\left(\Lambda(t_i,\mathbf{S}_{ij}),\sum\limits_{q=1}^L
        \widehat{\beta}_{iq}\psi_q(\mathbf{S}_{ij})\right)\,,j=1,\ldots,M$.
     \end{itemize}
     \item[(iv)] Set $\widehat{V}_0=(\widehat{V}_{11}+\ldots+\widehat{V}_{1M})/M$.
\end{enumerate}

To illustrate the regression algorithm above, we consider a Bermudan
put option on the maximum of two underlying assets $S_1$ and $S_2$
modeled by the \textsf{UOU} bridge copula with interest rate $r=0.05$,
zero dividend yield, strike price $K=100$, time to maturity $T=1.0$,
initial spot $S_0=100$, and $N=10$ exercise times distributed evenly
in the time interval $[0,T]$. Here we use the same choice of the
model parameters as in the previous numerical example.
Clearly, this example can easily be extended to the
case with three or more assets.

\begin{table}
  \centering
  \caption{\label{table3} Comparison of price estimates for Bermudan and European put options
  on the maximum of two assets modeled by
  the bridge copula multivariate \textsf{UOU} model with different values of the
  correlation coefficient~$\theta$. A sample standard error is given after $\pm$ sign.}
  \begin{tabular}{r@{~~~}r@{\,$\pm$\,}l@{~~~}r@{\,$\pm$\,}l}
      $\theta$ & \multicolumn{2}{c}{Bermudan} & \multicolumn{2}{c}{European} \\
\hline
      $-0.75$ & 3.336 & 0.002 & 1.831 & 0.002  \\
      $ 0   $ & 7.106 & 0.003 & 5.680 & 0.003 \\
      $ 0.75$ & 11.580 & 0.004 & 10.798 & 0.005 \\
\hline
  \end{tabular}
\end{table}

In the regression algorithm the basis functions are the power functions
$1,$ $S_1,$ $S_2,$ $S_1^2,$ $S_2^2,$ $S_1 S_2,$ and payoff function
$\Lambda(S_1,S_2)=\left(K-\max(S_1,S_2)\right)^+.$ We apply the
regression method with $M=100\,000$ independent paths. To calculate
standard error we replicate the random estimator $100$ times. Table
\ref{table3} shows numerical results for differing values of the
correlation coefficient $\theta$. For comparison's sake we also
calculate the values of the European put option on the maximum of
underlyings by the Monte Carlo method with $10^7$ sample paths.

\section*{Conclusion}
In this paper we have constructed a new multi-asset pricing
model, whose marginal (single-asset price) processes are nonlinear (local
volatility smile) regular diffusions from a recently developed \textsf{UOU}
family of probability conserving martingale models \cite{CM06,CM08b}. The multivariate model is based
on a bridge copula method, where a normal distribution function couples the underlying univariate Ornstein-Uhlenbeck bridges and consequently
forms a multivariate asset price process with built-in correlations.
Such an approach preserves the solvability of the model, hence the multivariate path density is available in closed form and
can be used for the calibration of the model to market prices. Extra flexibility of the multivariate model is provided by
different variations of the bridge sampling method. We are also able to sample multivariate paths from their exact distribution.
Moreover, the proposed exact bridge simulation algorithm runs faster than any approximation scheme (e.g., the Euler method).

To illustrate the financial applications of our model, we succeeded in calibrating the model to single-asset equity option
market prices as well as in calibrating the multi-asset price correlation matrix to historical asset prices. We also succeeded in pricing multi-asset
Asian and Bermudan options by using a path integral Monte Carlo (MC) approach and an MC regression method,
respectively. The preliminary calculations presented in this
paper pave the way to further applications of derivatives pricing
under such multi-asset state dependent diffusion models. The
numerical efficiency of the path integral MC approach used in this
paper can be further improved by employing quasi-MC methods. These methods have
already been successfully used for another nonlinear model, the
so-called Bessel-\textsf{K} model \cite{CM06,CM08b}, and are also
applicable to the \textsf{UOU} local volatility smile model.

\section*{Acknowledgements}
The authors acknowledge the support of the Natural Sciences and
Engineering Research Council of Canada (NSERC) for discovery
research grants as well as SHARCNET (the Shared Hierarchical
Academic Research Computing Network) in providing support for a
research chair in financial mathematics and a graduate scholarship.
SHARCNET, CFI and OIT are
also acknowledged in support of a New Opportunities Award for a
computer cluster equipment grant.



\end{document}